\documentclass[a4paper]{jpconf}
\pdfoutput=1

\usepackage{epsfig}
\usepackage{graphicx}
\usepackage{graphics}
\usepackage{url}
\usepackage{iopams} 

\newcommand\be{\begin{equation}}
\newcommand\ba{\begin{eqnarray}}
\newcommand\ee{\end{equation}}
\newcommand\ea{\end{eqnarray}}

\newcommand{\salto}[1]{\left[\,#1\,\right]^{}_{p}}

\newcommand{\met}{\mbox{g}}

\newcommand{\singu}{{\mbox{\tiny S}}}
\newcommand{\regu}{{\mbox{\tiny R}}}
\newcommand{\Hor}{{\mbox{\tiny H}}}

\begin{document}

\title{Simulations of Extreme-Mass-Ratio Inspirals Using Pseudospectral Methods}

\author{Priscilla Ca\~nizares, Carlos F. Sopuerta}

\address{Institut de Ci\`encies de l'Espai (CSIC-IEEC), Facultat de Ci\`encies, 
Campus UAB, Torre C5 parells, E-08193 Bellaterra, Spain}

\ead{pcm@ieec.uab.es, sopuerta@ieec.uab.es}

\begin{abstract}
Extreme-mass-ratio inspirals (EMRIs), stellar-mass compact objects (SCOs) inspiralling
into a massive black hole, are one of the main sources of gravitational waves expected 
for the Laser Interferometer Space Antenna (LISA).  To extract the EMRI signals from 
the expected LISA data stream, which will also contain the instrumental noise as well 
as other signals, we need very accurate theoretical templates of the gravitational 
waves that they produce.  In order to construct those templates we need to account for the 
gravitational backreaction, that is, how the gravitational field of the SCO affects its 
own trajectory.  In general relativity, the backreaction can be described in terms of a 
local {\em self-force}, and the foundations to compute it have been laid recently.  
Due to its complexity, some parts of the calculation of the self-force have to be 
performed numerically.  Here, we report on an ongoing effort towards the computation of 
the self-force based on time-domain multi-grid pseudospectral methods.
\end{abstract}

\section{Introduction}
Extreme-Mass-Ratio Inspirals (EMRIs) are one of the main source of gravitational waves for the planned Laser Interferometer Space Antenna (LISA)~\cite{LISA}. EMRIs are binary systems composed by a stellar compact object (SCO), like a neutron star or a stellar black hole, with masses in the range $m = 1-10^2 M_{\odot}$, inspiralling into a massive black hole (MBH), with masses in the range $M= 10^4-10^7 M_{\odot}$, located at galactic centers.  Then, the mass ratios involved are $\mu=m/M \sim 10^{-7}-10^{-3}$.  During the inspiral phase the system is driven by the emission of gravitational radiation, and hence there is loss of energy and angular momentum that makes the
orbit shrink until the SCO plunges into the MBH.  It is expected that LISA will be able to detect $10-10^3$ EMRI$/yr$ up to distances with $z \lesssim 1$~\cite{Gair:2004iv,Hopman:2006xn} (see also~\cite{AmaroSeoane:2007aw}). In order to extract the signals produced by these systems, which will be buried in instrumental noise and the gravitational wave foreground (produced by compact binaries in the LISA band), and extract relevant physical information from them, we need to have an {\em a priori} theoretical knowledge of the gravitational waveforms with a high degree of precision.  While techniques for constructing templates good enough for detection are getting ready, methods to build templates good enough for extraction of physical information are not yet fully developed. The main difficulty being that one requires a more precise treatment of the self-gravity of the SCO and its impact on the gravitational waveform.

Frequency-domain approaches provide very accurate results in the computation of quasinormal modes and frequencies and also in the computation of the self-force for EMRIs with moderate eccentricities. However, the frequency domain approach has more difficulties when dealing with EMRIs with highly eccentric orbits, which are of interest for LISA, since one has to sum over a large number of modes to obtain a good accuracy, and convergence may be an issue. 
On the other hand, time-domain approaches are not affected much by the eccentricity of the orbit and may be
more efficient for the case of high-eccentricity EMRIs.  However, in the time domain numerical approach one has to deal with different physical scales (both spatial and temporal) present in the problem due to extreme mass ratios.  Specifically,  we  have to computationally resolve large wavelength scales (comparable to the massive black hole) and also the scales in the vicinity of the SCO, which are crucial for evaluating the self-force.  This means, in practice, that we have to resolve numerically large scales and small scales inside the same computational domain.  Here we introduce a new time-domain scheme towards the computation of the self-force based on the PseudoSpectral Collocation (PSC) method (see, e.g.~\cite{Boyd}).  A remarkable property of this new scheme is that it avoids having to introduce in the problem a small spatial scale associated with the presence of the SCO (see, e.g.~\cite{Sopuerta:2005rd}).  This is done by using multiple domains and locating the particle in the interface bewteen two of them.  Therefore, we just
need the resolution to describe the field near the particle, but not the particle itself, which makes the
computation more efficient.  Here, we describe how to apply this technique to the computation of the self-force
on a charged scalar particle in circular orbits around a non-rotating black hole and show some results of the 
numerical implementation. 

\section{Study of a Simplified Model}
We study a simplified model of an EMRI consisting of a particle with charge $q$, associated with a scalar field $\Phi(x^{\mu})$, orbiting a non-rotating MBH. In this model, the particle generates the scalar field, which in turn influences the particle trajectory (this effect can be described by a local force, called the {\em self-force}).  Therefore, this model contains all the ingredients of the gravitational case (see details in~\cite{Mino:1997nk,Quinn:1997am}), where the particle motion is influenced by its own gravitational field instead of a scalar field. 

In the scalar field model the spacetime geometry is not dynamical. In our problem it has to describe the geometry of the MBH and therefore will be given by the Schwarzschild black hole metric:
\begin{equation}
ds^2=f(-dt^2+dr^\ast{}^{2})+r^2d\Omega^2\,,~~d\Omega^2=d\theta^2+\sin^2\theta d\varphi^2\,, \label{schmetric}
\end{equation}
where $(x^{\mu}) =(t,r,\theta,\varphi)$ are the Schwarzschild coordinates, $f=1-{2M}/{r}$, $M$ is the MBH mass, and  $r^{\ast} = r + 2M \ln\left(r/2M-1\right)$ is the {\em tortoise} coordinate.
The scalar field equation is a wave-like equation with a singular source term due to the energy density of the particle's scalar charge~(see, e.g.~\cite{Poisson:2003nc})
\begin{equation}
\met^{\alpha \beta}\nabla_{\alpha}\nabla_{\beta}\Phi(x)=-4\pi \rho = -4\pi q\int_{\gamma}d \tau\, \delta^{}_4(x,{z}(\tau))  \,, \label{geo}
\end{equation}
where $x^{\mu} = z^{\mu}(\tau)$ is the trajectory of the particle (without loss of generality, the orbital plane is taken to be the equatorial plane); $\nabla_{\mu}$ denotes the associated canonical connection; $\tau$ denotes proper time along the particle timelike worldline $\gamma$, and $\delta^{}_4(x,{z}(\tau))$ is the invariant Dirac functional in Schwarzschild spacetime~(see, e.g.~\cite{Poisson:2003nc}) and describes the singular structure of the source term.  The equations of motion for the particle, obtained from energy-momentum conservation, are
\begin{equation}
m\frac{du^{\mu}}{d\tau} = F^{\mu} = q\, (\met^{\mu\nu} + u^{\mu}u^{\nu})\nabla^{}_{\nu}\Phi\,, ~~
u^{\mu} = \frac{dz^{\mu}}{d\tau} \,, \label{particlemotion}
\end{equation}
where $m$ is the particle mass and $u^{\mu}$ its 4-velocity.  The force $F^{\mu}$ diverges on the particle worldline, $\gamma$, due to the singularity that carries the particle field, and therefore it must be regularized~\cite{Quinn:2000wa}. An analysis of the solutions of (\ref{geo}) and (\ref{particlemotion}) reveals (see~\cite{Poisson:2003nc} and references therein) that  the field, $\Phi$, can be split into two parts~\cite{Detweiler:2002mi}: $\Phi= \Phi ^{\regu} + \Phi^{\singu}$.  Where $\Phi^{\singu}$ is the singular part, which contains the singular structure of the field and satisfies the same field equation, and $\Phi ^{\regu}$ is the  regular part, which satisfies an homogeneous 
wave equation:
\begin{equation}
\met^{\alpha\beta}\nabla_{\alpha}\nabla_{\beta}\Phi^{\singu} =-4\pi \rho\, ,~~~~~\met^{\alpha\beta}\nabla_{\alpha}\nabla_{\beta}\Phi^{\regu} = 0 \,. \label{eqsingular}
\end{equation}
The regular field is solely responsible of the deviation of the particle from geodesic motion around the MBH. Hence,  the {\em self-force} which acts on the particle is $F^{\regu}_{\mu}=  q (\met^{\mu\nu} + u^{\mu}u^{\nu})\nabla^{}_{\nu}\Phi^{\regu}$. 

In order to solve the field equation~(\ref{geo}) we can make use of the spherical symmetry of the Schwartzchild metric and decompose $\Phi$ into scalar spherical harmonics. The equations for the different harmonic coefficients, $\Phi^{m}_{\ell}(t,r)$, decouple and satisfy 1+1 wave-like equations of the form
\begin{eqnarray}
\left\lbrace  -\frac{\partial^2}{\partial t^2} + \frac{\partial^2}{\partial r^{\ast}{}^{2}} -V^{}_{\ell}(r) \right\rbrace(r\Phi^{m}_{\ell})= f\,S^{m}_{\ell}\delta (r-r^{}_{p}(t))\,, 
\label{master}
\end{eqnarray}
where $V^{}_{\ell}(r)$ is the Regge-Wheeler potential for scalar fields on the 
Schwarzschild geometry and $S^{m}_{\ell}$ is the harmonic coefficient of the singular source term generated
by the particle, which is given by:
\begin{eqnarray}
V^{}_{\ell}(r) = f \left[ \frac{\ell(\ell+1)}{r^2}+\frac{2M}{r^3}\right]\,,~~~S^{m}_{\ell} = -\frac{4\pi qf(r^{}_{p})}{r^{}_{p}u^t}\,
\bar{Y}^{m}_{\ell}(\theta,\varphi^{}_{p})\,,
\label{source}
\end{eqnarray}
where an overbar denotes complex conjugation.  The solution of (\ref{master}) is finite and continuous at the particle location, but it is not be differentiable in the sense that the evaluation of the radial derivative of the scalar field from the left and from the right of the particle yields different values, that is, there is a jump [see equation~(\ref{jump}) bellow].  Nevertheless, one can see that the contribution to the self-force of each harmonic 
is finite, it is the sum over $\ell$ that becomes infinite.  Taking advantage of this fact, a method to extract
the singular part of the self-force mode by mode was developed, the so-called {\em mode-sum} regularization scheme~\cite{Barack:1999wf,Barack:2001gx,Barack:2002mha}.   Given the multipolar decomposition of the gradient of the scalar field
\begin{equation}
\Phi_{\alpha}(x^{\mu})=\sum_{\ell=0}^{\infty}\Phi^{\ell}_{\alpha}(x^{\mu})\,,~~\textrm{where}~~
\Phi^{\ell}_{\alpha}(x^{\mu}) = \nabla^{}_{\alpha}\sum_{m=-\ell}^{\ell}\Phi^{ m}_{\ell}(t,r)\,Y^{m}_{\ell}(\theta,\phi) \,,\label{l_retarded}\,
\end{equation}
each $\Phi^{\ell}_{\alpha}$ is finite at the particle location.  Using the mode-sum scheme, the regularized part of the gradient of the scalar field can be written as follows
\begin{equation}
\Phi^{\regu}_{\alpha}({z}^{\mu}(\tau))= \mathop{\lim}\limits_{x^{\mu}\to {z}^{\mu}(\tau)}\sum_{\ell=0}^{\infty} \left(\Phi^\ell_{\alpha}(x^{\mu})-{\Phi^{\singu,\ell}_{\alpha} }(x^{\mu})\right) \,,
\end{equation}
and the multipoles of the singular part, $\Phi^{\singu,\ell}_{\alpha}$, are known in a neighborhood of the particle worldline and are given by~\cite{Barack:2002mha}:
\begin{eqnarray}
\lim_{x^{\mu}\to z^{\mu}(\tau)} \Phi^{\singu,\ell}_{\alpha}  = q\left[ \left(\ell+\frac{1}{2}\right) A^{}_{\alpha} + B^{}_{\alpha} + \frac{C^{}_{\alpha}}{\ell +\frac{1}{2}} - \frac{2\sqrt{2}D_{\alpha}}{(2\ell-1)(2\ell+3)}+... \right]\,, 
\label{l_sing}
\end{eqnarray}
where $A_{\alpha}$, $B_{\alpha}$, $C_{\alpha}$, $D _{\alpha}$, \ldots\, are called the regularization parameters~\cite{Barack:2001gx,Barack:2002mha,Detweiler:2002gi}.  They are independent of $\ell$, 
but depend on the particle trajectory.  The singular part corresponds to the first three terms which lead to quadratic, linear, and logarithmical divergences. 
The remaining terms form a convergent series that does not contribute to the self-force (each of them).   
Since the only non-vanishing component of the regularization coefficients is the
radial one, this is the only component of the self-force that is actually singular, and hence the only one to be regularized.
 
\section{Description of the Simulations}
We begin by defining the spatial domain where we want to obtain our solution:  $r^{\ast}\in (-\infty,+\infty)$, where $r^{\ast}\rightarrow-\infty$ corresponds to the horizon location ($r=2M$), while $r^{\ast}\rightarrow+\infty$ corresponds to spatial infinity. For computational reasons, we have to take a {\em truncated} radial domain (we could avoid this by compactifying the radial domain):
 $r^{\ast}\in \Omega=[r^\ast_\Hor,r^\ast_\infty] $, and hence we need to prescribe outgoing boundary conditions at the end points of this domain:
\begin{eqnarray}
\partial^{}_{t}(r\,\Phi^{m}_{\ell} (t,r^{\ast}_{\Hor})) -\partial^{}_{r^\ast}( r\,\Phi^{m}_{\ell}(t,r^{\ast}_{\Hor}))= 0\, \label{boundary}
,~~~\partial^{}_{t}(r\,\Phi^{m}_{\ell} (t,r^{\ast}_{_\infty})) +\partial^{}_{r^\ast}( r\,\Phi^{m}_{\ell}(t,r^{\ast}_{_\infty}))= 0\,.
\end{eqnarray}
On the other hand, in order to numerically compute the solutions, it is necessary to discretize our equations in time and space.  We use the PSC for the spatial discretization, and the time discretization is performed by using a Runge-Kutta algorithm. Then, the harmonic components of the scalar field are approximated by an expansion in Chebyshev polynomials, 
\begin{equation}
\Phi_{\ell,N}^{m}(t,r^{\ast}) = \sum_{n=0}^{N} a_{n}(t)T_{n}(X(r^{\ast}))\,,~~~ X\in[-1,1]\,,
\end{equation}
and the coefficients $a_{n}(t)$ are determined, in terms of a system of ordinary differential equations (ODEs), by forcing the equation~(\ref{master}) to be satisfied at a given set of collocation points $\{X_k\}^{}_{k=0,1,\ldots,N}$.  For our calculations we took a {\em Lobatto-Chebyshev} grid, which includes the boundary points as nodes of the grid. 
The system of ODEs is solved by using a Runge-Kutta algorithm, typically a RK4 scheme.

The singular source term~(\ref{source}) in our equation~(\ref{master}) would in principle spoil the exponential
convergence of the PSC method.  In order to avoid this, and also in order to avoid a {\em small scale} associated with the presence of the particle, we use a multi-domain scheme and locate the particle
in the interface between two such domains.  More specifically, we divide  $\Omega$ into $D$ sub-domains $\Omega =\cup_i[r_i^{\ast\, L},r_i^{\ast\, R} ]$ ($i=1,...,D$), with $r^{\ast\,R}_{i-1}=r^{\ast\,L}_{i}$, and evolve the equation (without the source term) in each sub-domain independently.  In doing so we need to communicate the different subdomains, in particular the subdomains to the left and to the right of the particle, by using appropriate boundary conditions.  These conditions can be derived from the equation~(\ref{master}), and can be written as junction 
conditions that describe the possible jumps in the solution ($\salto{\lambda} = \mathop{\lim }\limits_{r^\ast \to r^\ast_{p}}\lambda^{}_{+}(t, r^\ast)-\mathop{\lim }\limits_{r^\ast \to r^\ast_{p}}\lambda^{}_{-}(t, r^\ast)$)
across the radial particle location (see~\cite{Sopuerta:2005gz} for details).  In the case of circular orbits
they are
\begin{eqnarray}
\salto{\Phi^{m}_{\ell} (t,r^{\ast}) } = 0\,,~~ \salto{\partial^{}_{t}\Phi^{m}_{\ell} (t,r^{\ast})} = 0\,,~~\salto{\partial^{}_{r^\ast}\Phi^{m}_{\ell} (t,r^{\ast})} = f S^{m}_{\ell}\,. \label{jump}
\end{eqnarray}

Apart from these techniques, we also use the Fast Fourier Transform in order to evaluate derivatives of the scalar field in a quick way (see, e.g.~\cite{Boyd}). Moreover, in order to reduce possible spurious high-frequency components of the numerical solution, we apply an exponential spectral filter to the solutions after every time step 
of the RK algorithm.  

\section{Summary of Results}
\begin{figure}[h] 
\begin{minipage}[c]{0.5\linewidth}
\centering
\includegraphics[width=7.2cm]{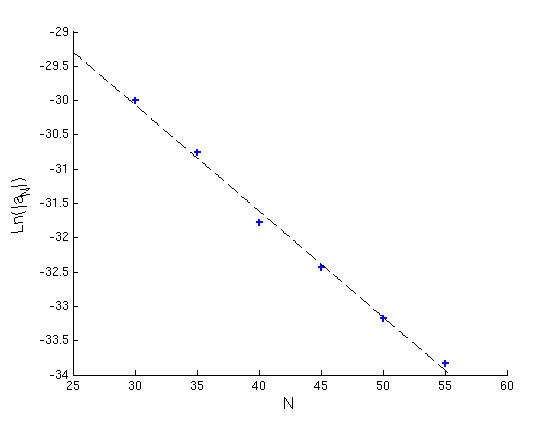}\\[1mm]
\includegraphics[width=7.2cm]{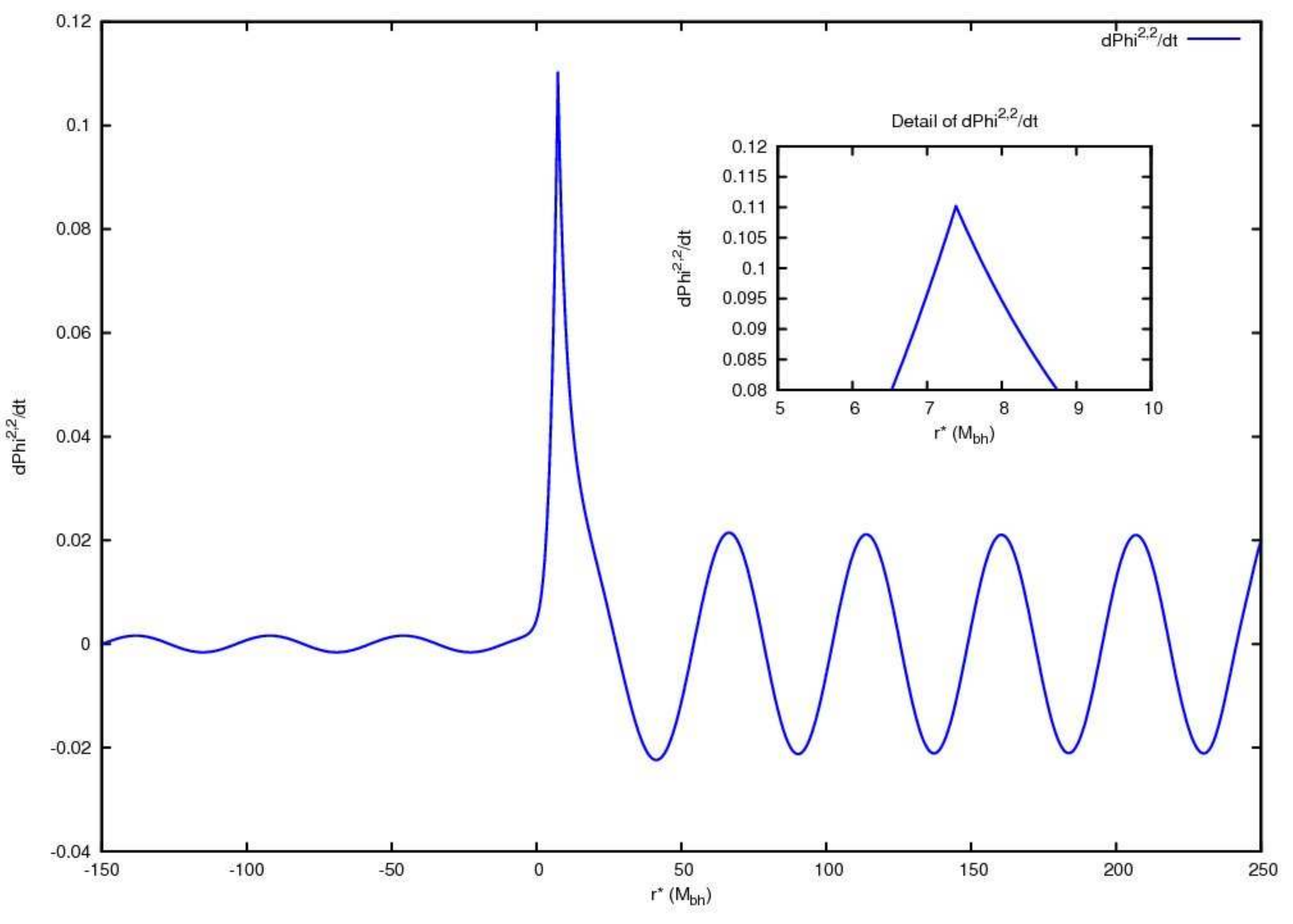}
\end{minipage}\hfill
\begin{minipage}[c]{0.5\linewidth}
\centering
\includegraphics[width=7.2cm]{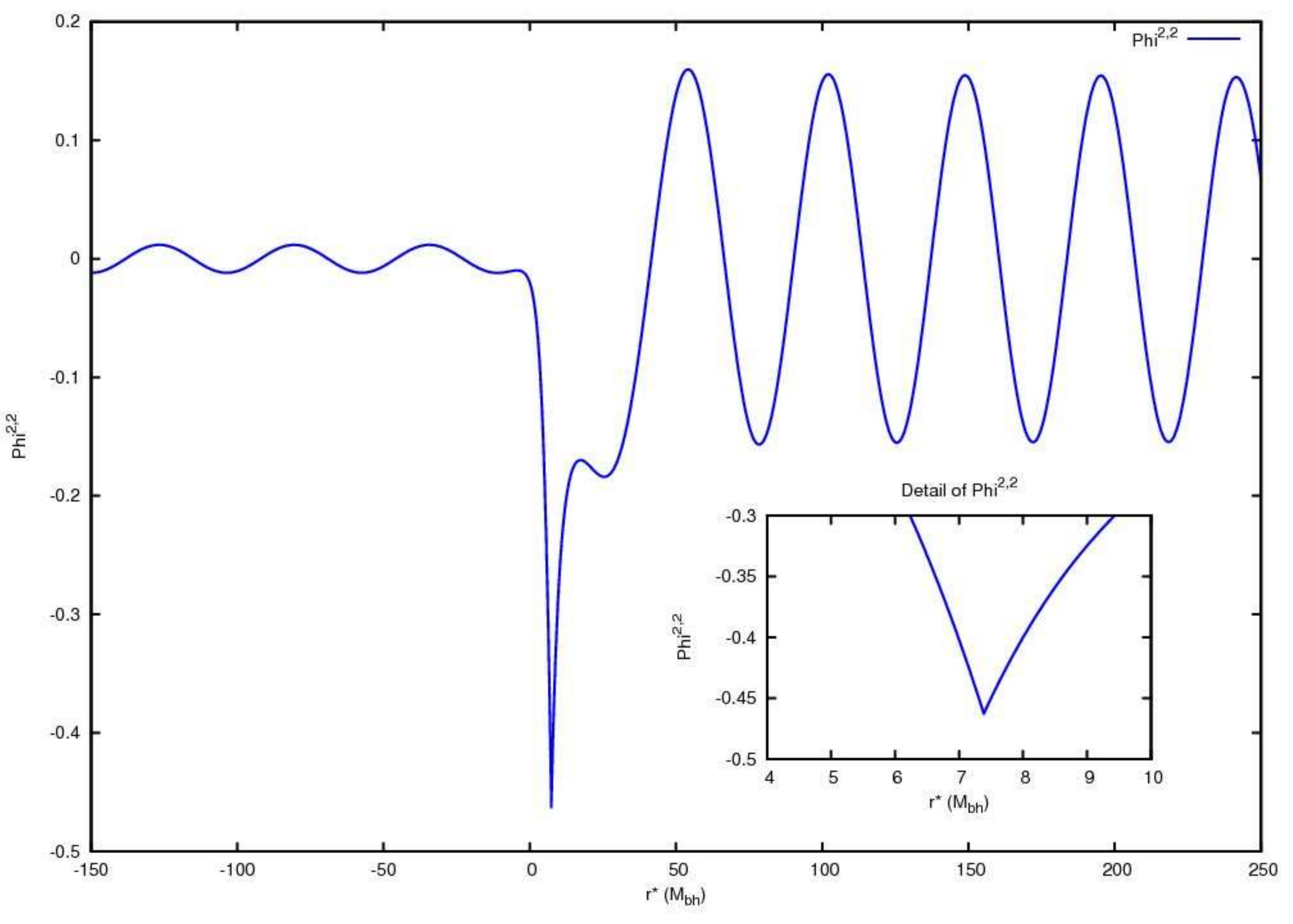} \\[10mm]
\includegraphics[width=7.2cm]{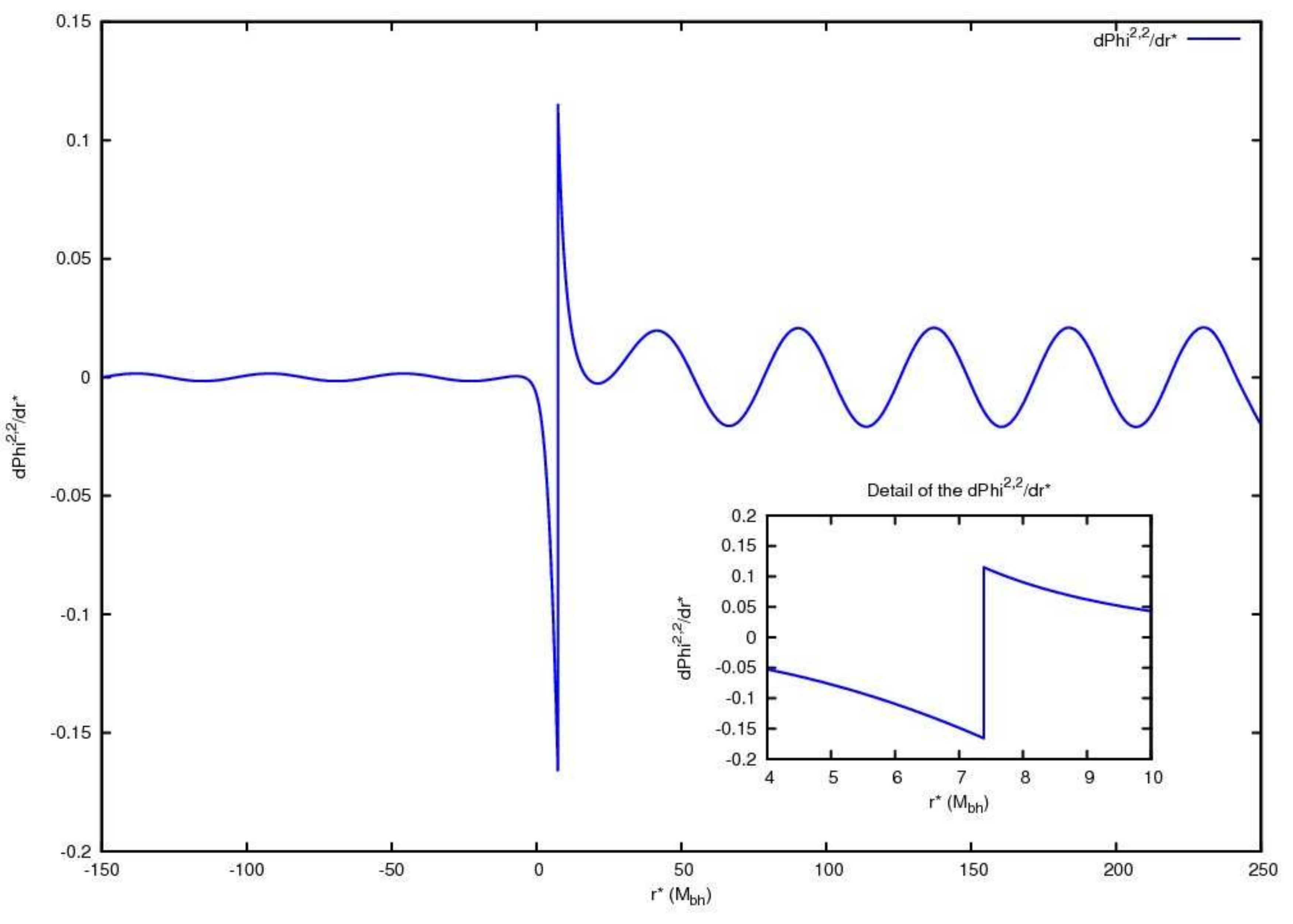}
\end{minipage}
\caption{ {\em Top left:} Estimation of the truncation error, $\ln |a_N|$, versus the number of collocation points, $N$. The points lie aroud a straight line with negative slope, showing that the expected exponential convergence is obtained.  The other figures show snapshots of the evolution of the $(2,2)$ mode (with $r_{p}=6M$) and the detailed structure near the particle.  They show the continuity of $\Phi_{\ell}^{m}$ ({\em top right}) and
$\partial^{}_{t}\Phi^{m}_{\ell}$ ({\em bottom left}), and the jump in $\partial^{}_{r^\ast}\Phi^{m}_{\ell}$ ({\em botton right}). \label{plots}}
\end{figure}

We have developed a numerical code that implements the previous ideas in order to compute the self-force
on a charged scalar particle in circular geodesics around a non-rotating MBH. 
In order to validite the code we have made a series of tests.  In particular, we have evolved a Gaussian 
pulse in flat space ($V_l=0$) and also in the Schwarzschild geometry ($V_l\neq0$), that is, we have evolved
the equation~(\ref{master}) without a particle.  We have done this using different numbers of collocation 
points, $N$, in order to check that we obtain the expected convergence of the PSC method for smooth solutions.
In Figure~\ref{plots} (top left) we show that indeed an estimation of the truncation error, $\ln|a_{N}|$, falls exponentially as the number of collocation points increases.

On the other hand, we have performed simulations including the charged particle in order to develop and test the communication between the subdomains, in particular between those around the particle where jumps in the solution can occur.  In practice, this communication is implemented by using a penalty method, in such a way that the jump conditions in equation~(\ref{jump}) are enforced in a dynamical way, by introducing new terms into the resulting ODEs that drive the solution towards the satisfaction of these conditions.  These new terms have to be callibrated against the number of collocation points and the size of the time step.   We have found that the penalty method, in combination with the use of a spectral filter, works very well and can produce accurate solutions to our equations.  In order to illustrate this fact we have included three plots in Figure~\ref{plots} corresponding to snapshots of the evolution of the $(\ell,m)=(2,2)$ mode at the last stable circular orbit ($r=6M$), which is probably the most demanding situation.  At the top right, we show a snapshot of the scalar field, $\Phi_{2}^{\;2}$, where it can be seen that the field it is continuous at the particle although it is not differentiable. At the bottom left we have a snapshot for the time derivative of the
scalar field, $\partial^{}_{t}\Phi_{2}^{\;2}$, which is also continuous (this only happens for the case of circular orbits~\cite{Sopuerta:2005rd}) but non-differentiable.  Finally, at the bottom right, we show a snapshot of the radial derivative of the field, $\partial^{}_{r^{\ast}}\Phi_{2}^{\;2}$, where we can see the ability our numerical techniques in resolving the jump accross the particle.  In the small box inside this figure we zoom in near the particle location to show how well the method can deal with the sharp jump and at the same time produce smooth waves.

\section{Conclusions and Future Work}
We have presented a new time-domain computational framework to compute the self-force for EMRIs. We have implemented this new method numerically for the case of a charged scalar particular moving in circular
geodesics around a non-rotating MBH.   This method is based on the PSC method.   One of the main ingredients of this technique is to consider a multi-domain grid and to locate the particle in the interface between two subdomains.  In this way, we solve equations with smooth solutions inside each subdomain, which preserves the powerful convergence properties of the PSC method and, at the same time, we avoid introducing a small scale related to the size of the SCO.
This in turn makes the algorithm quite efficient as compared to other treatments of the particle.
The effects in the solution due to the presence of the particle source term appear in our method through the imposition of a set of jump conditions.  Here, we have presented tests of the convergence of the numerical code we have built and we have also shown the ability of the code to resolve the field near the particle, including the jump in the radial derivative, and at the same time produce smooth waves.  Calculations of the self-force on the particle are underway and will be presented elsewhere~\cite{CanizaresandSopuerta}.

In this work, we have restricted ourselves to the case of circular orbits, where the particle is always at the same radial position.  In order to extend our calculations to generic eccentric orbits we need to introduce a method to deal with a moving particle, either by moving our domains or by changing the coordinate system so that the particle is always located at the same node.  This is a project presently under development.  Given that the astrophysically interesting EMRIs involve spinning MBHs, another important task for the future is to study how to implement all these techniques for the case in which the MBH is represented by the Kerr solution.

\ack
We would like to thank Leor Barack, Jos\'e Luis Jaramillo, Eric Poisson, for helpful discussions. 
CFS acknowledges support from the Ram\'on y Cajal Programme of the
Ministry of Education and Science of Spain and by a Marie Curie
International Reintegration Grant (MIRG-CT-2007-205005/PHY) within the
7th European Community Framework Programme.   

\section*{References}

\providecommand{\newblock}{}

\end{document}